\documentclass[useAMS,usenatbib]{mn2e}

\usepackage{natbib, aas_macros}

\usepackage{graphicx}

\title[The growth of massive stars via stellar collisions in ensemble star clusters]{The growth of massive stars via stellar collisions in ensemble star clusters}

\author[M. S. Fujii and S. Portegies Zwart]
{M. S. Fujii$^{1}$
\thanks{E-mail: fujii@strw.leidenuniv.nl(MSF); spz@strw.leidenuniv.nl(SPZ)} 
and S. Portegies Zwart$^{1}$\footnotemark[1]\\
$^{1}$Leiden Observatory, Leiden University, NL-2300RA Leiden, The Netherlands}
\begin{document}

\date{Accepted 1988 December 15. Received 1988 December 14; in original form 1988 October 11}

\pagerange{\pageref{firstpage}--\pageref{lastpage}} \pubyear{2002}

\maketitle

\label{firstpage}

\begin{abstract}

Recent simulations and observations suggest that star clusters form
via the assembling of smaller sub-clusters. Because of their short 
relaxation time, sub-clusters experience core collapse much earlier 
than virialized solo-clusters, which have similar properties of the 
merger remnant of the assembling clusters. As a consequence it 
seems that the assembling clusters result in 
efficient multiple collisions of stars in the cluster core.
We performed a series of $N$-body simulations of ensemble and 
solitary clusters including stellar collisions and found that the 
efficiency of multiple collisions between stars are suppressed
if sub-clusters assemble after they experience core collapse
individually. In this case, sub-clusters form their own multiple 
collision stars which experienced a few collisions, but they 
fail to collide with each other after their host sub-clusters assemble. 
The multiple collision stars scatter each other and escape, and 
furthermore the central density of the remnant clusters had already 
been depleted for the stars to experience more collisions. 
On the other hand, if sub-clusters assemble before they experience core 
collapse, the multiple collisions of stars proceed efficiently in the 
remnant cluster, and the collision products are more massive than 
virialized solo-clusters and comparable in mass to cold solo-clusters.   

\end{abstract}


\section{Introduction}
Young dense star clusters observed in the Milky Way and the Large 
Magellanic Cloud (LMC), e.g., 
R136 \citep{1998ApJ...493..180M,2010MNRAS.408..731C}, 
NGC 3603 \citep{2006AJ....132..253S,2008ApJ...675.1319H}, 
Westerlund 1 
\citep{2005A&A...434..949C,2008A&A...478..137B,2011MNRAS.412.2469G} 
and 2 \citep{2007A&A...466..137A,2007A&A...463..981R}, are 
good samples for understanding the formation mechanism of dense 
star clusters. They are massive ($\sim 10^5M_{\odot}$) and dense 
($>10^4M_{\odot}{\rm pc}^{-3}$), and seem to be approaching 
(or might have experienced) core collapse although they are 
young ($<4$Myr) \citep{2003MNRAS.338...85M}. For example, in R136 in 
the LMC, its high core density
($>5\times10^4 M_{\odot}{\rm pc}^{-3}$) \citep{2003MNRAS.338...85M}
and the existence of high-velocity stars (runaway stars) escaping
from the cluster \citep{2007ASPC..367..629B,2010ApJ...715L..74E,
2011A&A...530L..14B,2011MNRAS.410..304G} suggest that it 
experienced core collapse \citep{2011Sci...334.1380F}.
If such a young massive cluster experiences core collapse, 
repeated collisions (so-called runaway collisions) of stars, and 
as a consequence the formation of very massive stars ($>100M_{\odot}$), 
are expected \citep{1999A&A...348..117P,2002ApJ...576..899P,
2011MNRAS.410.2799M,2011MNRAS.413.1810B}.
Such very massive stars formed through multiple stellar collisions 
could result in the formation of intermediate-mass black holes (IMBHs)
\citep{2001ApJ...562L..19E}.

The formation of IMBHs in dense star clusters via 
multiple collisions has been studied
using $N$-body simulations \citep{1999A&A...348..117P,
2004Natur.428..724P,2004ApJ...604..632G,2006MNRAS.368..141F}, and 
the results suggest that IMBHs with $10^2-10^3M_{\odot}$
could be formed in such dense clusters.  
Including stellar evolution, however, a high 
mass-loss rate due to the stellar wind of massive stars prevents the 
growth of the massive stars \citep{2007ApJ...659.1576B,2009A&A...497..255G}. 
A very high collision rate is required for such very massive stars
to overcome the copious mass-loss and nevertheless leads to the 
formation of an IMBH \citep{2008ApJ...686.1082F}. 

There are some mechanisms to enhance the growth rate of the 
very massive stars, but the most important factor is the moment of 
core collapse, $t_{\rm cc}$. This short but high density phase is 
necessary for the cluster to become collisionally dominated, which is 
critical for the collision rate of stars in the cluster. 
Earlier collapse times assist an efficient mass 
accumulation because stars can start multiple collisions before 
the cluster starts to lose massive stars via stellar evolution.

The core-collapse time is determined by the relaxation time of the
virialized star cluster, which with a Salpeter-type mass function is
about 20\% of the half-mass relaxation time, $t_{\rm rh}$ 
\citep{2002ApJ...576..899P,2003gmbp.book.....H}. 
Massive clusters are unlikely to
reach core collapse before the end of main-sequence lifetime of their
most massive stars, which for $> 40M_{\odot}$ is $\sim 3$ Myr. 
These clusters can
still reach core collapse before the most massive stars leave the
main-sequence if they are born kinematically relatively cold. A
sub-virial cluster evolves faster dynamically than a cluster that is
born in vitial equilibrium; mass segregation in sub-virial clusters
also proceeds on a shorter time scale \citep{2009ApJ...700L..99A}. Mass
segregation as well as core collapse proceed on the same --dynamical--
timescale, and for sub-virial clusters also the increase of the core
density proceeds on a shorter time scale than for virialized clusters.
The accelerated dynamical evolution of sub-virial clusters enables an
efficient mass-growth via multiple collisions of stars.

Mass segregation causes the massive stars to sink to the cluster
center, and consequently to pile up in the cluster core. In the core
these star find each other and initiate a collision runaway
\citep{1999A&A...348..117P}. The consequent mass-growth due to stellar
collisions can be quite efficient, in particular if the massive stars
are concentrated in the core. Allowing a cluster to be born with some
degree of mass segregation also makes the collision runaway more
efficient \citep{2008ApJ...682.1195A,2012ApJ...752...43G}, much in the 
same way as sub-virial initial conditions reduces the time of
mass-segregation, which again leads to an enhanced collision rate
\citep{2009ApJ...700L..99A}.

Fractal initial conditions and assembling sub-cluster models also
result in an early dynamical evolution similarly to sub-virial initial 
conditions \citep{1972A&A....21..255A,
2007ApJ...655L..45M,2009MNRAS.400..657M,2009ApJ...700L..99A, 
2011ApJ...732...16Y,
2011MNRAS.416..383S,2012ApJ...753...85F}.  The short relaxation time
of sub-clumps compared to initially massive single clusters causes
early mass segregation and core collapse.  The memory of such early
dynamical evolution is conserved in the merger remnant
\citep[][hereafter Paper 1]{2007ApJ...655L..45M,2012ApJ...753...85F},
the formation of star clusters by assembling them seems to be an
effective way for efficient multiple collisions of stars in young star
clusters.

In paper 1, we found that the formation scenario of young dense 
star clusters via mergers of ensemble sub-clusters can 
successfully explain the mature characteristics of young massive 
star clusters such as R136 in 30 Dor region. The age of R136 
is only 2--3 Myr, but it shows dynamically mature characteristics, 
such as mass segregation, a high core density, and a wealth of high 
velocity escaping stars \citep{2003MNRAS.338...85M,2007ASPC..367..629B,
2010ApJ...715L..74E,2011A&A...530L..14B,2011MNRAS.410..304G}. 
However, the relaxation time of R136 obtained from its current mass 
and radius is $\sim 100$ Myr
\citep{2003MNRAS.338...85M}, which is too long to have
reached core collapse at its current age. In paper 1, 
we performed a series of $N$-body simulations of ensemble clusters 
and demonstrated that ``ensemble''-cluster models can reproduce
observations such as the core density, the fraction of 
high-velocity escapers, and the distribution of massive stars which 
experienced collisions, but ``solo''-cluster models, which are 
initially spherical and virialized, fail to reproduce these observations. 
Furthermore, these 
characteristics of the ensemble models are also consistent with the 
characteristics of other massive young clusters like 
R136 in the LMC and 
NGC 3603 in the Milky Way \citep{2010MNRAS.408..731C}. 

If young dense clusters formed via assembling sub-clusters 
and have experienced core collapse, it is expected that repeating 
collisions can lead to the formation of very massive stars and 
possibly even IMBHs. In the observed young dense 
clusters, however, there is no evidence of IMBHs, but
some very massive stars 
with an initial mass of 100--300 $M_{\odot}$ are observed 
\citep{2006AJ....132..253S,
2011A&A...530L..14B, 2009Ap&SS.324..321M,
2011MNRAS.412.2469G,2011MNRAS.416..501R}.

In this paper, we perform a series of $N$-body simulations of 
solo and ensemble star clusters and demonstrate that the 
growth of very massive stars through multiple collisions is 
mediated by star cluster complexes. Our simulations show that 
the quick dynamical evolution of 
ensemble clusters does not always result in the formation of 
extremely-high-mass stars.
When the assembling of clusters proceeds after each sub-cluster 
experiences core collapse (``late-assembling'' case), 
multiple-collision stars that form in each sub-cluster fail 
to coalesce to an extremely massive star, 
but leads to the formation of several very massive stars. 
Some of these very massive stars can escape from the cluster as 
high-velocity stars due to the three-body or binary-binary 
encounters. When the sub-clusters assemble before they experience 
core-collapse (``early-assembling'' case), the collision rate
is enhanced and the assembled cluster forms an extremely massive star 
of $\sim 1000 M_{\odot}$.

\section{Method and Initial Conditions}
We performed a series of $N$-body simulations of solo clusters 
and ensemble clusters, that merge to a single cluster with a mass
equal to solo clusters. For the ensemble of sub-clusters, we 
adopted two models. A:
a King model \citep{1966AJ.....71...64K} with 
a dimensionless concentration parameter, $W_0$, of 2 and the 
total mass $M_{\rm cl}=6300M_{\odot}$, and B: a King model 
with $W_0=5$ and $M_{\rm cl}=2.5\times 10^4 M_{\odot}$.
The half-mass radii, $r_{\rm h}$, of these models are 0.092 and 
0.22 pc, and the numbers of particles, $N$, are 2048 (2k) and 
8192 (8k), respectively. 
The core density is the same for both models 
($\rho_{\rm c} \simeq 2 \times 10^6 M_{\odot}\,\mathrm{pc}^{-3}$). 
We assumed a Salpeter initial mass function (IMF) 
\citep{1955ApJ...121..161S} between 1 and 100 
$M_{\odot}$. We call these models 2kw2 and 8kw5.  

\begin{figure*}
\begin{center}
\includegraphics[width=70mm]{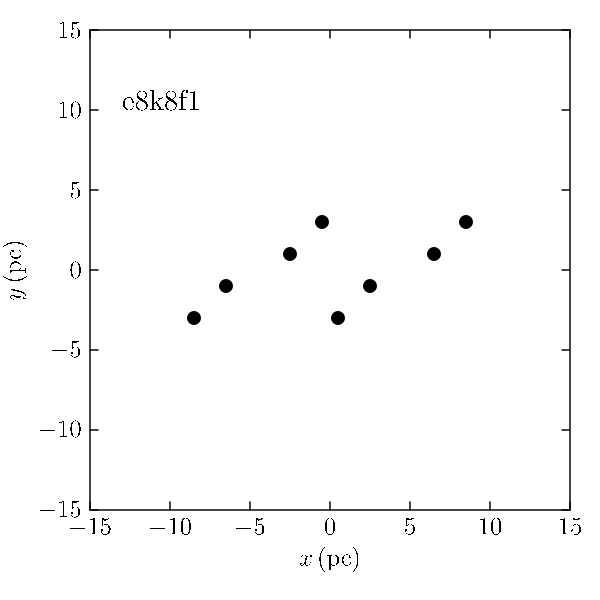}
\includegraphics[width=70mm]{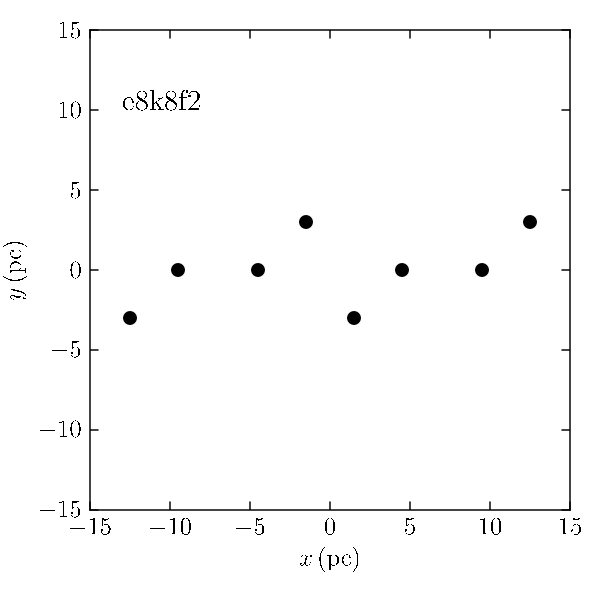}
\caption{Initial position of ensemble models, e8k8f1 (right) and e8k8f2 (left). We mimicked filamentary star forming regions. \label{fig:init_pos}}
\end{center}
\end{figure*}

We distribute 4 or 8 of these sub-clusters in two different 
initial configurations: spherical or
filamentary. The former model stems from clumpy star
formation in giant molecular clouds, and the latter is motivated 
by star formation in a filamentary gas distribution or shocked region
of colliding gas in the spiral arms of a galactic disk. 
The clumpy star formation is initiated by observations of Westerlund 1
\citep{2011MNRAS.412.2469G} and R136 \citep{2012ApJ...754L..37S} and 
simulations \citep{2011MNRAS.410.2339B,2011IAUS..270..483S}.
For the spherical models, we adopted 4 or 8 of models 2kw2 as 
sub-clusters, and distributed them randomly in a volume with a radius 
of $r_{\rm max}$ and with zero velocity.
We varied $r_{\rm max}$ between 1 and 6 pc.
For the filamentary models, we initialized 8 individual 8kw5 
model sub-clusters.
We initialized these sub-clusters with 
two different initial mean separations (models e8k8f1 and e8k8f2),
but with zero velocity. 
The initial positions of the sub-clusters for these models are 
illustrated in Figure \ref{fig:init_pos}.  
All runs are summarized in table \ref{tb:model}.

\begin{figure}
\begin{center}
\includegraphics[width=80mm]{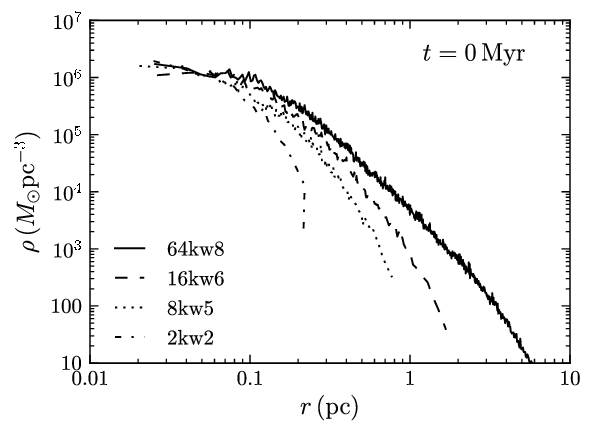}
\caption{Initial density profiles of single clusters\label{fig:density_init}}
\end{center}
\end{figure}

For the solo models, we adopted two more initial conditions with  
$M_{\rm cl}$ of $5.1 \times 10^4 M_{\odot}$ and $2.0 \times 10^5 M_{\odot}$.
With the same mass function, these models have 
16384 (16k) and 65536 (64k) stars and are initialized
using King models with $W_0=6$ and 8, respectively.
In order to obtain the same core density as that of sub-clusters,
their half-mass radii are 0.32 and 1.0 pc.
We call these models as 16kw6 and 64kw8. In Table \ref{tb:model_cl}
we summarize the initial conditions, and we present 
their initial density profiles in Figure \ref{fig:density_init}.
We performed additional simulations of sub-virial (cold)
initial conditions for 16kw6, and an extra set of simulations 
in which we reduced the kinetic energy 
(velocity of each particle) to two-thirds and 10\% of the 
virialized velocity.
We call these models as s16k-cool and s16k-cold, respectively
(see table \ref{tb:model}).

The $N$-body simulations are performed using the sixth-order Hermite
scheme with individual timesteps with an accuracy parameter
$\eta=$0.15--0.3 \citep{2008NewA...13..498N}.  We adopted the accuracy
parameter to balance speed and accuracy, and the energy error was $<
0.1$ \% for all runs.  Our code does not include special treatment for
binaries, but the sixth-order Hermite scheme can handle hard binaries
formed in our simulations (see section 2 in Paper 1).  We took into
account collisions of stars with a sticky-sphere approach and mass
loss due to the stellar wind for stars with $>100M_{\odot}$ with a
rate of $5.0\times 10^{-7} (m/M_{\odot}) M_{\odot}$yr$^{-1}$
\citep{2009ApJ...695.1421F}, which is similar to that obtained in
\citep{2007ApJ...659.1576B,2012A&A...538A..75P}.  We neglected the
mass-loss from stars with $<100M_{\odot}$ because it does not affect
the results on the short timescale of our simulations ($<5$Myr).  The
stellar radii are taken from the zero-age main-sequence for solar
metallicity and the radii follow \citet{2000MNRAS.315..543H} for stars
with up to $100M_{\odot}$. For more massive stars, we extrapolate the
result around $100M_{\odot}$ and the result is similar to SeBa stellar
evolution software
\citep{1996A&A...309..179P,1998A&A...332..173P,2012A&A...546A..70T}.
The details of the code are described in \citet{2009ApJ...695.1421F}.

\begin{table*}
\begin{center}
\caption{Models of single clusters\label{tb:model_cl}}
\begin{tabular}{cccccccccc}\hline \hline
Model & $N$ & $M_{\rm cl}$  & $W_0$ & $r_{\rm h}$ & $\rho_{\rm c}$& $\sigma$ & $t_{\rm rh}$ & $t_{\rm rc}$ &$M_{\rm core}/M_{\rm cl}$ \\ 
 &  &  ($M_{\odot}$) &  &  (pc) & ($M_{\odot}$pc$^{-3}$) & (km/s) & (Myr) &  (Myr) \\\hline
2kw2  & 2048 & $6.3\times 10^3$ & 2 & 0.097  & $1.7\times 10^6$ & 11 & 0.30 & 0.58 & 0.28 \\
8kw5  & 8192 & $2.5\times 10^4$ & 5  & 0.22  & $1.7\times 10^6$ & 15 & 1.9 & 0.92 & 0.15 \\
16kw6 & 16384 & $5.1\times 10^4$ & 6  & 0.32 & $1.7\times 10^6$ & 17  & 4.4 & 1.1 & 0.12\\
64kw8 & 65536 & $2.0\times 10^5$ & 8  & 1.0  & $1.6\times 10^6$ & 19 & 44 & 1.8 & 0.053 \\ \hline
\end{tabular}
\medskip
\\
$\sigma$ is the velocity dispersion.
\end{center}
\end{table*}

\begin{table*}
\begin{center}
\caption{Runs\label{tb:model}}
\begin{tabular}{cccccc}\hline
Model & $N_{\rm cl}$ & geometry & $\langle d_{\rm min}\rangle$ (pc) & (sub-)cluster & $N_{\rm run}$ \\ \hline\hline
e2k4r3  & 4 & spherical  &  2.5 & 2k2w & 3\\ 
e2k4r6  & 4 & spherical  &  5.1 & 2k2w & 1\\ \hline
e2k8r1  & 8 & spherical   & 0.51  & 2k2w & 2\\
e2k8r3  & 8 & spherical   & 1.3  & 2k2w & 1\\
e2k8r5  & 8 & spherical   &  2.8 & 2k2w & 2\\
e2k8r6  & 8 & spherical   &  3.3 & 2k2w & 2\\ \hline
e8k8f1  & 8 & filamentary & 2.8 & 8kw5 & 1\\
e8k8f2  & 8 & filamentary & 4.2 & 8kw5 & 1\\ \hline \hline
s2k     & 1 & - & - & 2kw3 & 7 \\  
s8k     & 1 & - & - & 8kw5 & 6\\
s16k    & 1 & - & - & 16kw6 & 6\\ 
s64k    & 1 & - & - & 64kw8 & 2\\ \hline
s16k-cool & 1 & - & - & 16kw6 & 2\\ 
s16k-cold & 1 & - & - & 16kw6 & 1\\ \hline
\end{tabular}
\medskip
\\
The models are named according to the following rules; ``e''  and ``s'' indicate ensemble and solo models, respectively. For ensemble models, following numbers indicate the number of particles of sub-clusters and the number of sub-clusters. The last part indicates the initial configuration of sub-clusters; ``r'' and the following number mean spherical and the value of the maximum radius, $r_{\rm max}$, ``f'' indicates filamentary initial configurations (see figure \ref{fig:init_pos} for the initial positions of sub-clusters in these models). For solo models, the number indicates the number of particles.
$\langle d_{\rm min}\rangle$ is the averaged distance to the nearest-neighbour sub-clusters, and $N_{\rm run}$ is the number of runs.
s16k-cool and s16k-cold are the same model as s16k, but 
the velocity of 67\% and 10\% of s16k, respectively.
\end{center}
\end{table*}

\section{Solo-cluster models}
\subsection{Virialized solo-cluster models}

\begin{figure}
\begin{center}
\includegraphics[width=84mm]{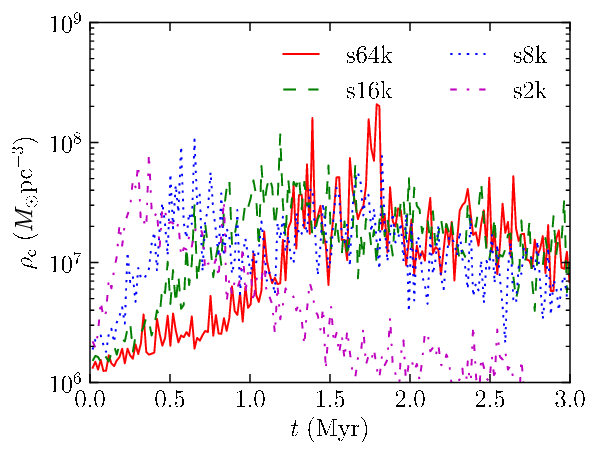}
\caption{Time evolution of the core densities for solo clusters. The results 
are averaged in order to reduce the run-to-run 
variations.\label{fig:cd}}
\end{center}
\end{figure}

We describe the results of the initially virialized 
solo-cluster models, which we will refer as the ``standard'' model.
In Figure \ref{fig:cd} we present the time evolution of the core 
density for models s2k, s8k, s16k, and s64k. 
The core densities are calculated using the method of 
\citet{1985ApJ...298...80C}.
We identify the moment when the cluster reaches the highest 
core density as the core-collapse time.
The core-collapse time measured from the simulations is 
$t_{\rm cc} = 0.29\pm 0.07$, $0.71\pm 0.11$, $1.2\pm 0.13$, and 
$1.8\pm 0.0$ Myr for models s2k, s8k, s16k, and s64k, respectively
(see also Table \ref{tb:results}). The core-collapse time is 
consistent with those obtained by previous simulations
\citep{2004ApJ...604..632G}, if we take into account the differences 
in the mass range of the mass function. As is demonstrated in 
\citet{2004ApJ...604..632G},
the core-collapse time scales with the central relaxation time 
\citep{2003gmbp.book.....H}:
\begin{eqnarray}
t_{\rm rc} = \frac{0.065 \sigma_{\rm c, 3D} ^3}{G^2 \langle m\rangle \rho_{\rm c} \ln \Lambda}.
\label{eq:tcrlx}
\end{eqnarray}
Here $G$, $\langle m \rangle$, $\sigma _{\rm c}$, and, $\rho_{\rm c}$
are the gravitational constant, the mean mass of stars, and the
central velocity dispersion and density, respectively. Here $\ln
\Lambda$ is the Coulomb logarithm.  In our simulations, $t_{\rm
  cc}/t_{\rm rc}\simeq 1$ for models s8k, s16k, and s64k, but $t_{\rm
  cc}/t_{\rm rc}\simeq 0.5$ for model s2k.  For model s2k, however,
$t_{\rm rh}$ is shorter than $t_{\rm rc}$ because the core radius
exceeds the half-mass radius. If we adopt a shorter relaxation time,
then $t_{\rm cc}/t_{\rm rh}\sim 1$ for all the models.  

\begin{figure*}
\begin{center}
\includegraphics[width=70mm]{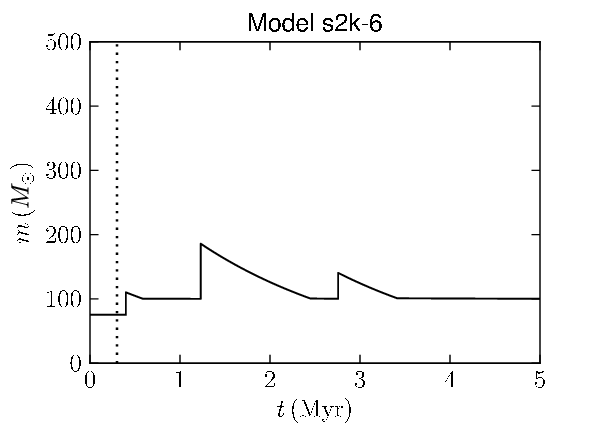}
\includegraphics[width=70mm]{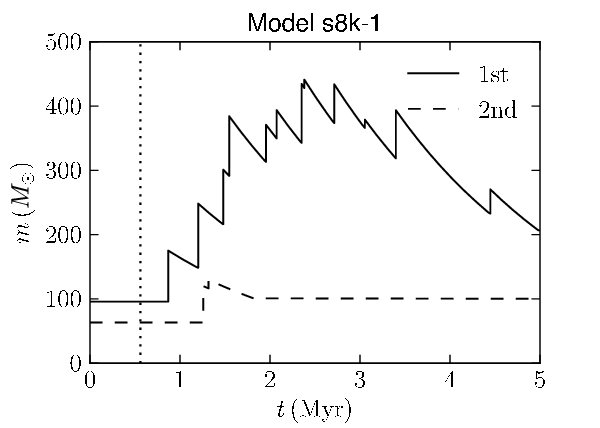}
\includegraphics[width=70mm]{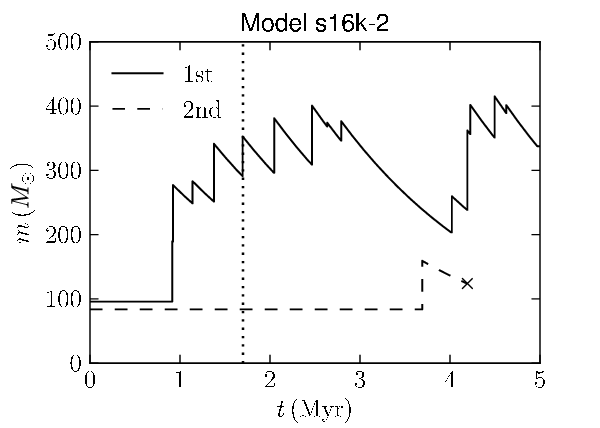}
\includegraphics[width=70mm]{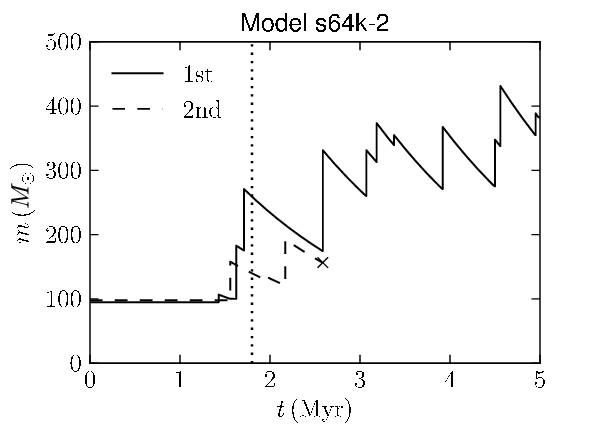}
\caption{Mass evolution of PCPs (solid curve) and SCPs (dashed curve) for models s2k, s8k, s16k, and 64k. Dotted line indicates the core-collapse time. Cross indicates the time when the star merged with more massive ones.\label{fig:m_his_single}}
\end{center}
\end{figure*}

The core collapse of the cluster initiates a collision runaway in the
cluster core \citep{1999A&A...348..117P}. In Figure
\ref{fig:m_his_single} we present the merger histories of the
multiple-collision stars in the solo-cluster simulations s2k, s8k,
s16k, and s64k. In each model, one primary collision product (PCP) per
cluster grows through repeated collisions of stars.  In model s2k the
mass-loss due to the stellar wind exceeds the mass-gain by the
collisions, and therefore the PCP has lost all gained mass by the end
of the simulation (5Myr).  PCPs grow up to the maximum mass $m_{\rm
  max}\sim 400 M_{\odot}$ via repeating collisions, but by the time it
explodes \citep[2--3 Myr][]{2007ApJ...659.1576B,2012A&A...538A..75P}
the star is $\sim 100M_{\odot}$. Here we define $m_{\rm max}$ as the
maximum mass of a star reached during its lifetime as a result of
collisions.

PCPs are not the only stars that experienced collisions. In models
s8k, s16k, and s64k, we find secondary collision products (SCPs). 
In most cases SCPs experience only one collision 
(sometimes a few collisions), but never 
grow as massive as PCPs, although SCPs sometimes exceed our adopted 
upper-limit to the IMF ($100M_{\odot}$). The SCPs end up merging with PCPs 
(see bottom right panel in Figure \ref{fig:m_his_single}) 
or just lose their mass by stellar evolution (see top right panel 
in Figure \ref{fig:m_his_single}). This result agrees with previous 
numerical simulations \citep{2006MNRAS.368..141F}.

We also find that the time when the PCPs reach their maximum mass 
$m_{\rm max}$, 
$t_{\rm max}$, is scaled by $t_{\rm rc}$, and that 
$t_{\rm max}/t_{\rm rc}=$  2.3, 2.2, 2.2, and 2.6 for models
s2k, s8k, s16k, and s64k, respectively.

\subsection{Cold solo-cluster models}

\begin{figure}
\begin{center}
\includegraphics[width=84mm]{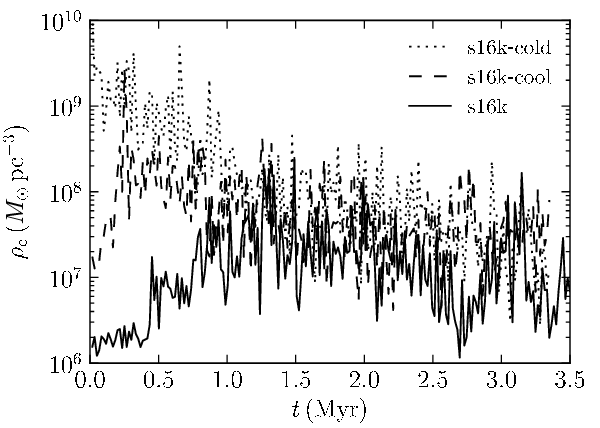}
\caption{Time evolution of the core density, $\rho_{\rm c}$, for models s16k, s16k-cool, and s16k-cold.\label{fig:cd_cold}}
\end{center}
\end{figure}

\begin{figure}[htbp]
\begin{center}
\includegraphics[width=84mm]{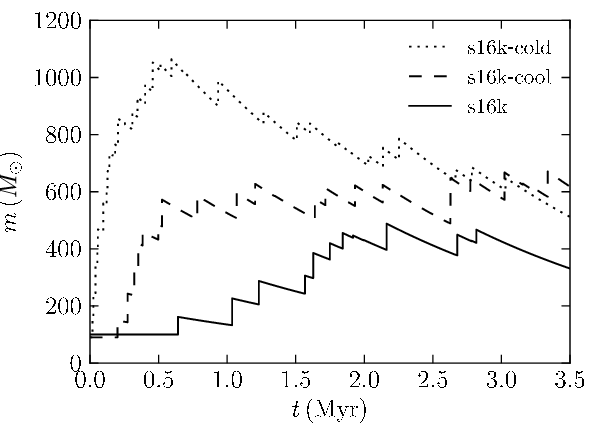}
\caption{Merger history of PCPs for models s16k, s16k-cool, and s16k-cold.\label{fig:m_his_cold}}
\end{center}
\end{figure}

Sub-virial (cold) initial conditions reach core collapse considerably 
earlier than virialized ones. Cold models have therefore been 
suggested to explain the dynamically advanced appearance of observed 
young star clusters \citep{2009ApJ...700L..99A}. 
In Figure \ref{fig:cd_cold} we present the core-density evolution of 
models s16k-cool and s16k-cold, which initially have 67\% and 10\% of 
the virialized temperature. These models reach core collapse much 
earlier than virialized models, and as a consequence multiple collisions 
start earlier and proceed at a higher collision rate.
In figure \ref{fig:m_his_cold} we present the mass evolution
of the PCPs for models s16k-cold and s16k-cool. 
Colder initial conditions result in a higher $m_{\rm max}$ of the 
PCPs. The high $m_{\rm max}$ is a result of the high collision 
rate, which is caused by the high density in the core 
(see Figure \ref{fig:cd_cold}). 

By the time the PCPs leave the main sequence (of $\sim 3$Myr), 
their masses have been reduced considerably due to stellar mass-loss,
which competes with the mass gain by collisions. 
In model s16k-cold, the PCP grows quickly in the beginning
of the simulation, but after 0.5 Myr the mass-loss rate due to the 
stellar wind becomes higher than the mass-growth by stellar collisions.  
In model s16k-cool, the PCP stops growing at $\sim$0.5 Myr because the 
mass-growth rate balances to the mass-loss rate, 
and then the PCP maintains its mass until the end of the simulation (3.5 Myr).
Although the final masses of the PCPs are comparable in both s16k-cool and 
s16k-cold models, $m_{\rm max}$ of model s16k-cold is twice as massive as
that of model s16k-cool.

\begin{figure}
\begin{center}
\includegraphics[width=84mm]{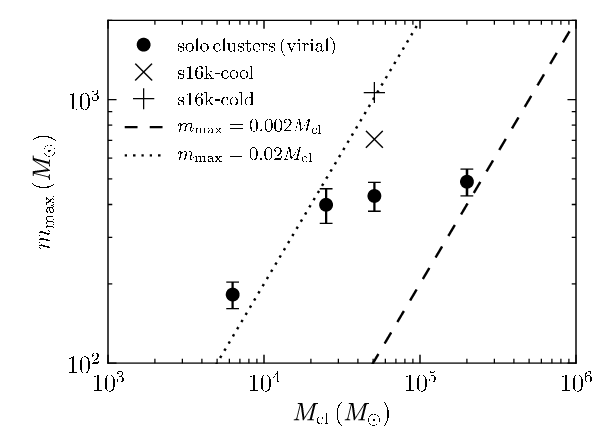}
\caption{The maximum mass of PCPs for solo models. Filled circles with error bars indicate models s2k, s8k, s16k, and s64k from left to right. Cross and plus indicate s16k-cool and s16k-cold, respectively. Since the error bar for model s16k-cool is smaller than the marker size, we do not plot the error bar. Dotted and dashed lines indicate $m_{\rm max} = 0.02M_{\rm cl}$ and  $m_{\rm max} = 0.002M_{\rm cl}$, respectively. \label{fig:m_max_single}}
\end{center}
\end{figure}

\begin{figure}
\begin{center}
\includegraphics[width=84mm]{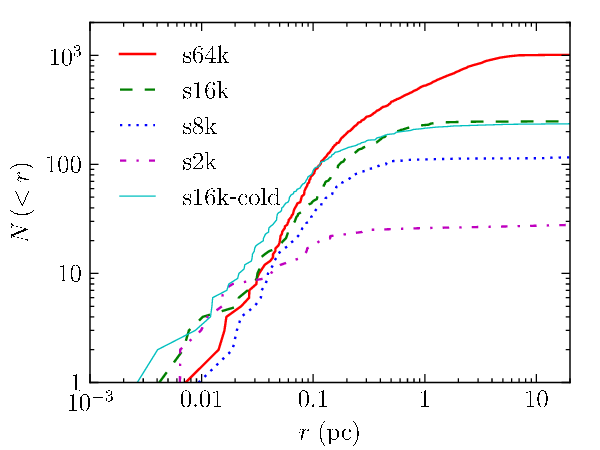}
\caption{Cumulative number distribution of stars with $m>50M_{\odot}$ at $t_{\rm max}$ for models s2k, s8k, s16k, s64k, and s16k-cold. \label{fig:nr_m50}}
\end{center}
\end{figure}

In Figure \ref{fig:m_max_single} we show $m_{\rm max}$ of the PCPs 
for all solo models. 
The maximum mass of the PCPs in models s8k, s16k, and s64k is quite 
similar ($\sim 400 M_{\odot}$) irrespective of $M_{\rm cl}$. 
One might expect that more massive clusters contain a larger number of 
massive stars and therefore a more massive cluster can form a more massive 
PCP. In our simulation, however, the 
number of stars which merged into the PCPs and the mean mass of the 
merged stars are quite similar among these models 
(see Table \ref{tb:results}).
By comparing models s8k, s16k, and s64k, their $\langle m_{\rm col} \rangle$
and $N_{\rm col}$ are quite similar even though their total cluster masses are
different. If the collisions selectively occur among the most
massive stars and the numbers of collisions are the same, larger clusters
should have a larger mean collision mass $\langle m_{\rm col} \rangle$ because 
larger clusters 
contain more massive stars. However, the number of massive stars does not
simply follow this relation. In figure \ref{fig:nr_m50} we plot the
cumulative number distribution of massive stars with $m>50M_{\odot}$ at 
the moment in which the mass of the PCP reaches $m_{\rm max}$.
The number of stars with $>50M_{\odot}$ within $\sim$0.05 pc are similar 
($\sim 20$) among models s8k, s16k, and s64k and slightly smaller for 
model s2k. In particular for the models s16k and s64k,
the distribution of massive stars preserves the initial distribution
in the outer part of the cluster because the half-mass relaxation time
exceeds $t_{\rm max}$.  The dynamical evolution in these models is driven 
on a timescale of $t_{\rm rc}$, and they have similar 
core properties: $t_{\rm rc}$, $M_{\rm core}$, and $\rho _{\rm c}$ 
(see Table \ref{tb:model_cl} and Figure \ref{fig:cd}).

In model s2k, on the other hand, $t_{\rm rc}\sim t_{\rm rh}$ and as a consequence
the dynamical evolution proceeds throughout the
entire cluster. Model s8k shows an evolution similar to that of 
model s2k: $m_{\rm max}/M_{\rm cl}$ for model s8k is as high as 
that of model s2k. For these models $t_{\rm rh}\sim 2$ Myr, which is 
sufficiently short for massive stars in the outer part of the cluster
to join the collisions in the core.
Similar to the model s2k, models s16k-cool and s16k-cold can
also gather massive stars from the entire cluster to the cluster center 
irrespective of their initial positions. In addition,
these sub-virial models achieve very 
high density (see Figure \ref{fig:cd_cold}), which enhances 
the collision rate. The massive stars in model s16k-cold are more 
concentrated towards the cluster center compared with model s16k
(see Figure \ref{fig:nr_m50}).

Even though for model s2k the PCP can accumulate stars from 
the entire cluster population of massive stars, their total number 
and mass still cannot compete with the population of massive stars 
in the more massive clusters.
In these latter models, the maximum mass of the PCP is limited
by the reservoir of massive stars, which manages to segregate to the 
core by the moment of the core collapse. A larger cluster mass 
therefore does not automatically lead to a massive PCP.
As seen in Figures \ref{fig:m_his_single} and \ref{fig:m_his_cold}, 
the mass evolution of the PCPs in models s2k, s8k, and s16k-cold 
show a clear peak in the middle of the simulation. In the later phase, 
when the collision rate decays, 
their mass-loss rate exceeds their mass-growth rate by stellar 
collisions. 
In models s16k and s64k, on the other hand, they have not exhausted
their reservoir of massive stars because their half-mass relaxation 
time is not shorter than the simulation time and therefore some of 
the massive stars still remain in the outer part of the clusters.

We empirically obtained a relation that $m_{\rm max} = 0.02M_{\rm cl}$ 
(dotted line in Figure \ref{fig:m_max_single}) for the low 
cluster-mass models ($M_{\rm cl} < 2\times 10^4M_{\odot}$) and 
the cold model.
For massive clusters, however, $m_{\rm max}$ is smaller than that
according to this 
relation. For the most massive cluster ($M_{\rm cl} = 2\times 10^5M_{\odot}$), 
$m_{\rm max}$ is consistent with the result presented by 
\citet{2002ApJ...576..899P}, $m_{\rm max} = 0.002M_{\rm cl}$ 
(dashed line in Figure \ref{fig:m_max_single}).

\begin{table*}
\begin{center}
\caption{Summary of the results.\label{tb:results}}
\begin{tabular}{ccccccccc}\hline
Model  & $m_{\rm max} (M_{\odot})$ & $t_{\rm max}$ (Myr)& $t_{\rm merge}$ (Myr)& $t_{\rm cc}$ (Myr) & $m_{\rm SCPs} (M_{\odot})$ & $\langle m_{\rm col}\rangle (M_{\odot})$ & $N_{\rm col}$ \\ \hline\hline
e2k4r3-1 & 287 & 0.45 & 0.2--0.87 & $0.29 \pm 0.07$ & 375 & 78.3 & 8 \\
e2k4r3-2 & 260 & 0.78 & 0.03--1.2 &  & 454 & 60.6 & 15 \\
e2k4r3-3  & 268 & 0.61 & 0.6--1.3 &  & 139 & 56.7 & 12  \\ 
e2k4r6  & 238 & 0.57 & 2.2--2.7 &   &  743 & 46.8 & 16 \\
\hline
e2k8r1-1    & 998 & 0.80 & 0.03--0.38 & $0.29 \pm 0.07$ & 0 & 44.2 & 45 \\
e2k8r1-2    & 667 & 1.35 & 0.03--0.32 &  & 160 & 69.4 & 22 \\
e2k8r3    & 530 & 0.86 & 1.0 --0.75 &  & 147 & 80.2 & 14 \\
e2k8r5-1  & 334 & 1.11 & 0.47--1.9 &  & 1192 & 57.0 & 25 \\
e2k8r5-2  & 486 & 0.78 & 0.03--2.0 &  & 651 &  61.2 & 19 \\
e2k8r6-1  & 245 & 0.59 & 0.77--2.4 &  & 1367 &  51.0 & 24   \\
e2k8r6-2  &  274 & 0.42 & 0.03--$>3$ &  & 970 & 45.5 & 21  \\ 
\hline
e8k8f1    & 1310 & 1.40  & 0.4--1.2 & $0.71\pm 0.11$ & 268 & 73.0 & 42 \\
e8k8f2    & 659  & 2.28 & 0.8--2.0 & & 995 &  88.4 & 33\\ 
\hline\hline
s2k       & $182 \pm 21$  & $1.3 \pm 0.6$ & - & $0.29 \pm 0.07$ & $16 \pm 40$  &53.3 & 4.6  \\
\hline
s8k     & $399\pm 60$ & $2.2\pm 0.2$ & - & $0.71\pm 0.11$ & $149 \pm 115$  & 63.2 & 11.3    \\ 
\hline
s16k    & $431\pm 54$ & $2.6\pm 0.9$ & - & $1.2\pm 0.13$ & $54 \pm 77$ & 65.8 & 13.2\\
\hline
s64k    & $488\pm 57$ & $4.4\pm 0.2$ &- & $1.8 \pm 0.0$ & 0 & 66.0 & 15.5 \\ 
\hline
s16k-cold    & 1064 & 0.59 & - & $<0.02$ & 0 &  46.6 & 40\\ 
s16k-cool   & $707\pm 36$ & $2.55\pm 0.75$ & - & $0.325\pm0.075$ & 0 & 51.1 & 28.5 \\ 
\hline
\end{tabular}
\end{center}
\end{table*}

\section{Ensemble-cluster models}

In section 3 we demonstrated that the results obtained from 
our solo-cluster models are consistent with previous numerical studies. 
In this section 
we present the results of ensemble-cluster models, in which 
sub-clusters assemble to finally form one single cluster. 
In ensemble-cluster models, sub-clusters collapse on a timescale
shorter than that for solo-clusters with the same total mass. 
Their further
evolution is dominated by the dynamical evolution of the sub-clusters 
before they merge. The conservation of the dynamical states through 
the mergers \citep{2009Ap&SS.324..277V} drives the further evolution 
of the cluster merger products. As a result, ensemble
clusters tend to experience core collapse considerably faster than 
solo clusters which have initially similar properties to those of the 
merger remnant of ensemble clusters. 
In paper 1 we already showed that the quicker dynamical evolution of 
ensemble clusters can explain the mature characteristics of 
young dense clusters such as R136 and NGC 3603. Here we use that
enhanced dynamical evolution to study the PCPs.
The early dynamical evolution of ensemble clusters is similar 
to that of cold solo-clusters. One might expect that ensemble 
clusters also result in the formation of massive PCPs, but 
we will show that the early evolution of 
ensemble clusters is more complicated.

\begin{figure*}
\begin{center}
\includegraphics[width=140mm]{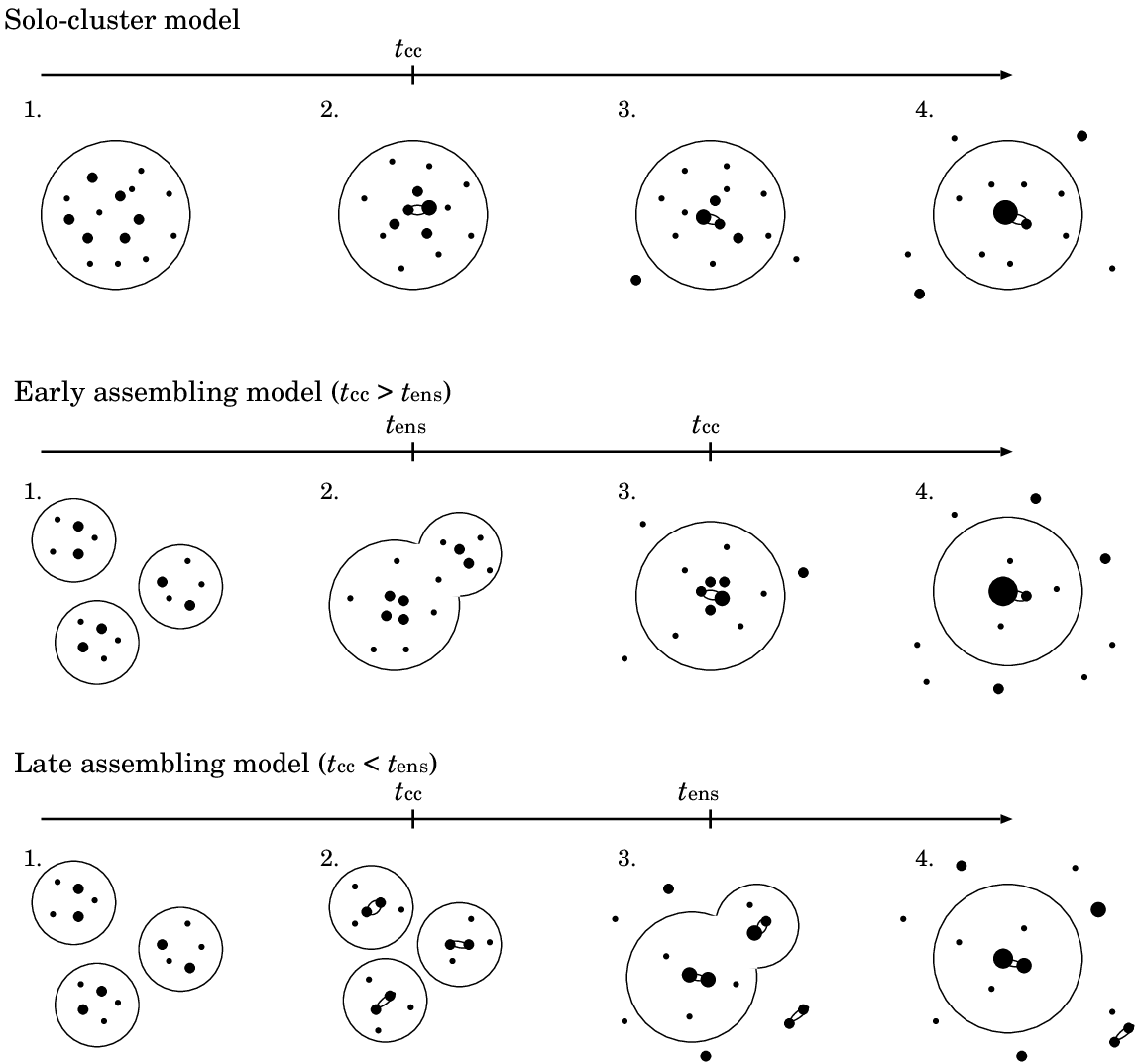}
\caption{Schematics picture of two typical assembling processes. Early assembling ($t_{\rm cc}>t_{\rm ens}$): Sub-clusters assemble before they experience core collapse. The merger remnant is more mass-segregated than solo clusters which initially have similar properties to the merger remnant because sub-clusters have a shorter relaxation time than the solo cluster. After their assembling, the remnant cluster collapses and a massive PCP forms. Late assembling ($t_{\rm cc}<t_{\rm ens}$): sub-clusters experience core collapse and form small PCPs before they assemble. After their assembling, however, the PCPs do not grow efficiently because most of them are scattered from the remnant cluster by binary-binary encounters.
\label{fig:merger}}
\end{center}
\end{figure*}

In Figure \ref{fig:merger}
we illustrate the schematic evolution of two typical evolutionary 
paths of ensemble clusters. We find that
the most important parameter for the evolution of ensemble clusters 
is the moment of assembling, $t_{\rm ens}$, compared to $t_{\rm cc}$ of 
sub-clusters. 
If $t_{\rm cc}>t_{\rm ens}$ (``early assembling''), the PCPs in the 
remnant cluster grow efficiently by stellar collisions because 
the short relaxation time of the sub-clusters drives 
mass-segregation and core collapse faster than solo clusters.
This evolution is similar to that of cold solo-clusters.

If $t_{\rm cc}<t_{\rm ens}$ (``late assembling''), each
sub-cluster experiences core collapse before they assemble and
form a PCP per individual sub-cluster. The mass of each PCP is limited 
by the sub-cluster mass as we described in section 3. 
After the assembling of two or more sub-clusters, 
the PCPs formed in the sub-clusters sink to the center of the remnant 
cluster and interact each other. Most of them, however, are 
scattered and ejected from the cluster because they tend to reside in
hard binaries with a massive companion. 
The PCPs tend to be in the hardest binaries with the
most massive stars when they formed in the
sub-clusters. 
In each binary-binary encounter following a sub-cluster merger, 
two PCPs may collide although they are also ejected without experiencing
a collision. 
Therefore, the majority of the PCP-binaries are scattered or ionized, 
and only one PCP-binary survives in the remnant cluster by the time the 
assembly is completed. The surviving PCP cannot continue to grow in 
mass because by that time the central density of the assembled clusters 
has been depleted due to the early dynamical evolution.

\subsection{Stellar collisions in ensemble clusters\label{sc_ensamble}}

\begin{figure*}
\begin{center}
\includegraphics[width=70mm]{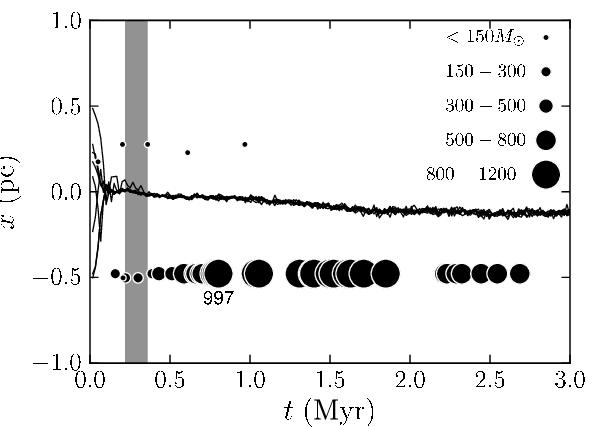}
\includegraphics[width=70mm]{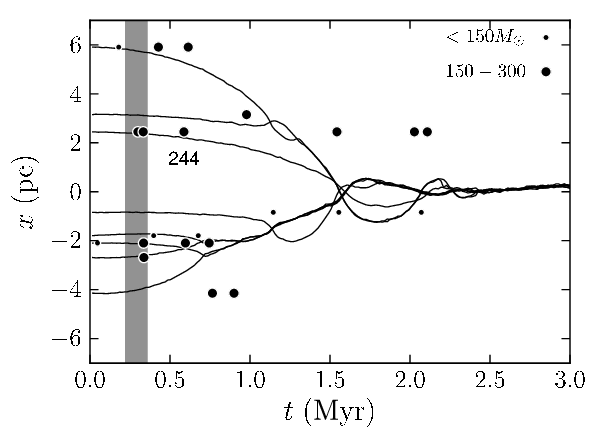}
\includegraphics[width=70mm]{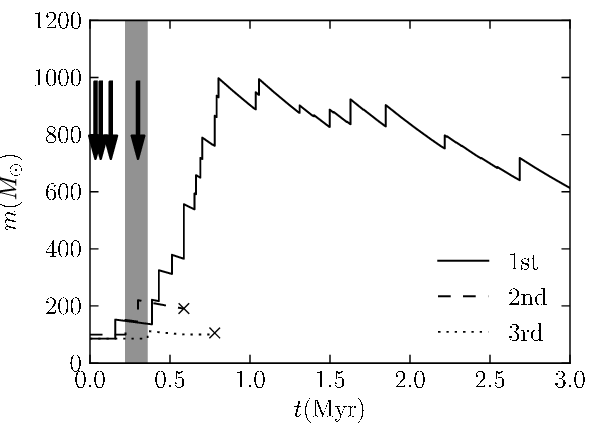}
\includegraphics[width=70mm]{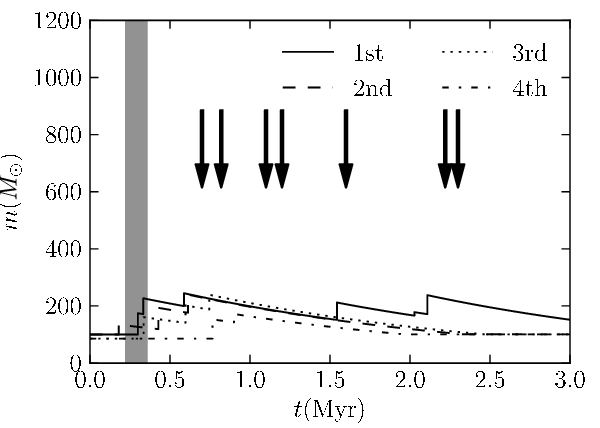}
\caption{Top: Time evolution of the separation between sub-clusters projected onto $x$-axis (full curves) and the collisions of PCPs (black dots) for models e2k8r1 (left) and e2k8r6-1 (right). The positions of the dots show the collision time and the the sub-cluster to which the star initially belongs.
Bottom: Mass evolution of PCPs and SCPs for models e2k8r1 (left) and e2k8r6-1 (right). Crosses indicate the time when the SCPs merged to PCPs. Arrows indicate the time when sub-clusters merged.
In all panels, the shaded region indicates the core-collapse time with error obtained from the simulations of isolated sub-clusters.  \label{fig:m_his_2k_8}}
\end{center}
\end{figure*}

\begin{figure*}
\begin{center}
\includegraphics[width=70mm]{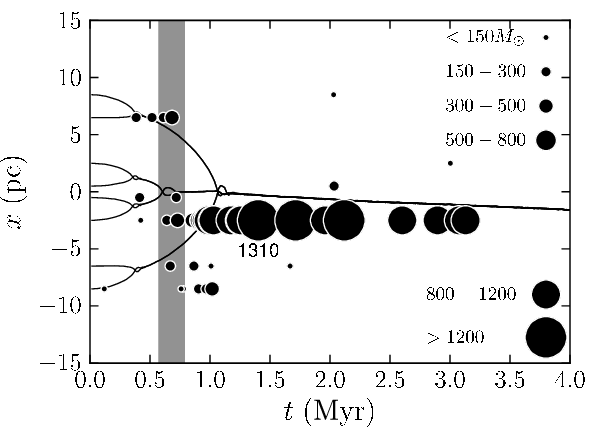}
\includegraphics[width=70mm]{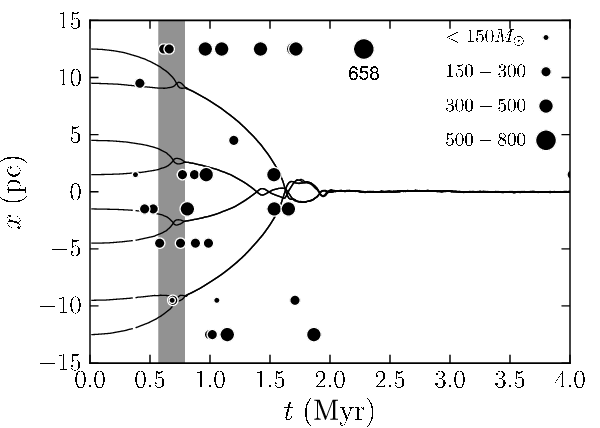}
\includegraphics[width=70mm]{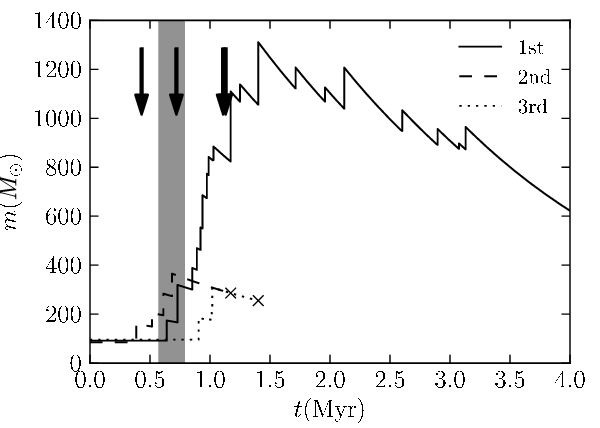}
\includegraphics[width=70mm]{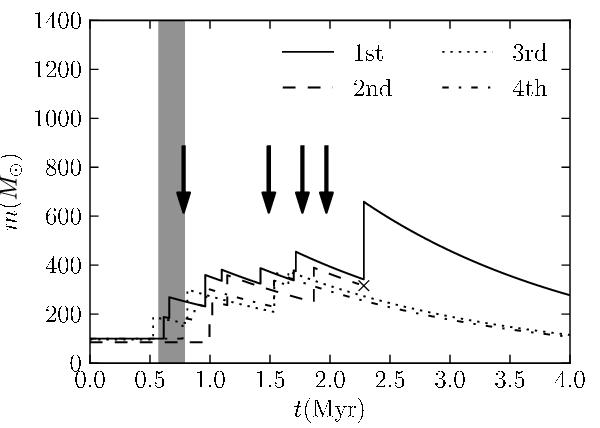}
\caption{Same as Figure \ref{fig:m_his_2k_8} but for models e8k8f1 (left) and e8k8f2 (right). \label{fig:m_his_8k_8}}
\end{center}
\end{figure*}

In Figures \ref{fig:m_his_2k_8} and \ref{fig:m_his_8k_8}, we present 
the mass evolution of PCPs and SCPs in ensemble clusters. The left and 
right panels show early and late assembling models, respectively. 
In early assembling models, one massive PCP per remnant cluster grows 
after the assembling of sub-clusters. Even though some of the sub-clusters 
start forming PCPs before assembling, the PCPs merge 
after the host sub-clusters merged. 
In late assembling models, on the other 
hand, each sub-cluster grows its own PCP, but most of them do not collide 
with each other even after the assembling of their host sub-clusters.

\begin{figure}
\begin{center}
\includegraphics[width=84mm]{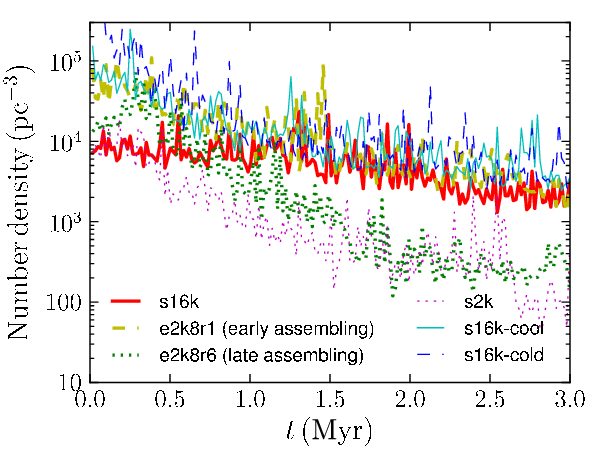}
\caption{Time evolution of maximum local number density (local densities of six nearest neighbours) for models s16k, s2k, s16k-cool, s16k-cold, e2k8r1 (early assembling), and e2k8r6 (late assembling). \label{fig:n_dens_merger}}
\end{center}
\end{figure}

We find the reason for the difference between early and late assembling 
cases in the density evolution of these clusters. 
In Figure \ref{fig:n_dens_merger} we show the time evolution of the 
maximum number densities for ensemble and solo clusters. Here we 
plot the maximum value of the local density, which is calculated 
using six nearest neighbours. (Note that the maximum local density 
does not trace the density of one individual sub-cluster.) 
In early assembling cases, the density increases on 
the core-collapse
timescale of the solo sub-cluster (model s2k), but the maximum 
density is higher than that of model s2k and rather comparable to 
those of the cold models (models s16k-cold, s16k-cool). The evolution 
after the core collapse is similar to that of the 
cold models. The density gradually decreases and eventually becomes 
comparable to that of virialized solo-clusters (model s16k). 

The density in late assembling cases also grows on the core-collapse 
timescale of the sub-clusters until a peak is reached at $\sim 0.5$ Myr. 
The density decreases as quickly as that of the solo sub-clusters 
(model s2k), which is different from early assembling cases. By the 
end of the simulations, the number density of the late assembling cases 
is an order of magnitude lower than in the early assembling
cases. The relatively low density prevents the growth of PCPs 
in the late assembled clusters. The effect of the difference in the 
density can be seen in the number 
of stellar collisions, $N_{\rm col}$ in Table \ref{tb:results}. 
In early assembling models (e2k8r1-1 and e8k8f1) and the cold solo model
(s16k-cold), $N_{\rm col}= 42\pm 2$ and $m_{\rm max}= 1100\pm 130$, but 
in late assembling models (e2k8r5, e2k8r6, and e8k8f2) $N_{\rm col}=24 
\pm 5$ and  $m_{\rm max}= 400\pm 150$. 

In late assembling models (for e2k8r5 and e2k8r6), the maximum mass 
of the PCPs is 200--400 $M_{\odot}$, but the mass of the PCP is similar
to those of multiple SCPs, which were PCPs in the sub-clusters. 
This feature is consistent with young dense
clusters such as R136 in the LMC, which contains
five $>100M_{\odot}$ mass stars
\citep{2010MNRAS.408..731C,2011A&A...530L..14B},
although there is no evidence of any
extremely massive stars with $\sim 1000 M_{\odot}$. 

\subsection{Maximum mass of PCPs in ensemble clusters}

\begin{figure}
\begin{center}
\includegraphics[width=84mm]{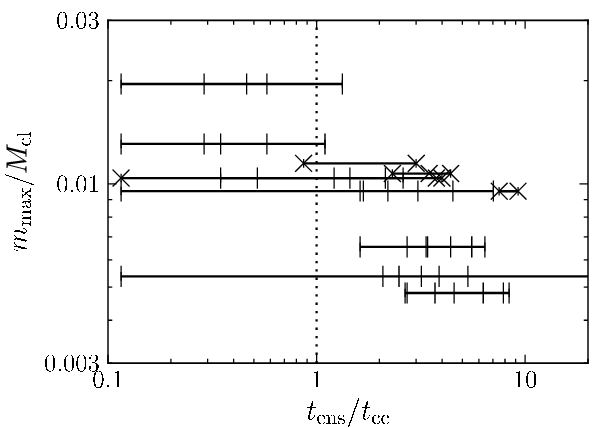}
\caption{The maximum mass of the PCPs scaled by the total mass of the 
ensemble clusters as a function of the assembling time scaled by the 
core-collapse time of the sub-clusters for all e2k8 and e2k4 models. 
Each horizontal line corresponds to one model. Vertical lines and
crosses indicate the individual 
merger time of sub-clusters for four and eight sub-cluster models, 
respectively. The dotted line indicates $t_{\rm ens}/t_{\rm cc} = 1$.
\label{fig:m_max_t}}
\end{center}
\end{figure}

As we show in section \ref{sc_ensamble}, early assembling of 
sub-clusters results in the formation of a PCP, while late assembling 
forms a less massive PCP and multiple SCPs as massive as the PCP. 
In Figure \ref{fig:m_max_t} we present the relation between 
$m_{\rm max}/M_{\rm cl}$ and $t_{\rm enc}/t_{\rm cc}$ of ensemble models,
where $t_{\rm cc}$ is the core-collapse time of the sub-clusters. 
Irrespective of the number
of sub-clusters, the maximum mass of the PCPs decreases as the 
assembling time is delayed.

\begin{figure*}
\begin{center}
\includegraphics[width=70mm]{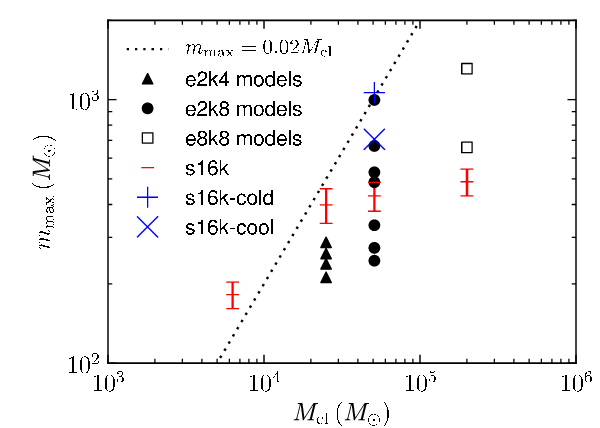}
\includegraphics[width=70mm]{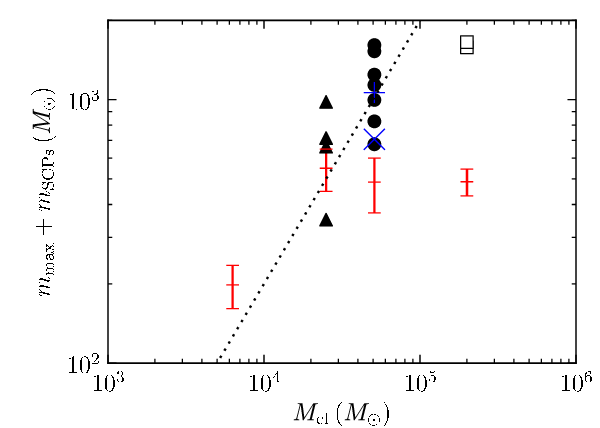}
\caption{The maximum mass of the PCP, $m_{\rm max}$, in the cluster (left) and the total mass of $m_{\rm max}$ and the sum of the maximum mass of the SCPs, $m_{\rm SCPs}$ (right). \label{fig:m_max}}
\end{center}
\end{figure*}

In the left panel of Figure \ref{fig:m_max}, we show the relation 
between $m_{\rm max}$ of the PCPs and $M_{\rm cl}$ for both
solo and ensemble clusters. (Note that for the solo clusters, the data is 
the same as that shown in Figure \ref{fig:m_max_single}). 
The PCP mass of early assembling models is higher than that of 
solo clusters with the same cluster mass and as massive as that of the cold 
model. In late assembling
models, the PCPs is almost as massive as those of the solo clusters
with the same mass.

The difference in the maximum mass of PCPs is understood if we 
take into account all the PCPs and SCPs in the cluster.
In the right panel of Figure \ref{fig:m_max}, we present the total
mass of all the PCPs and SCPs in the cluster. The total masses are 
roughly located on the relation that 
$m_{\rm max} = 0.02 M_{\rm cl}$. 
This result suggests that the potential maximum mass of the PCPs
is 2\% of the cluster mass, although the value depends on 
the initial mass function and the mass-loss rate due to the stellar
wind. The total mass of the SCPs is 
summarized in Table \ref{tb:results} as $m_{\rm SCPs}$. These SCPs 
fail to merge with the most massive PCP and their mass will be lost 
from the cluster by escape or stellar evolution.

\begin{figure}
\begin{center}
\includegraphics[width=84mm]{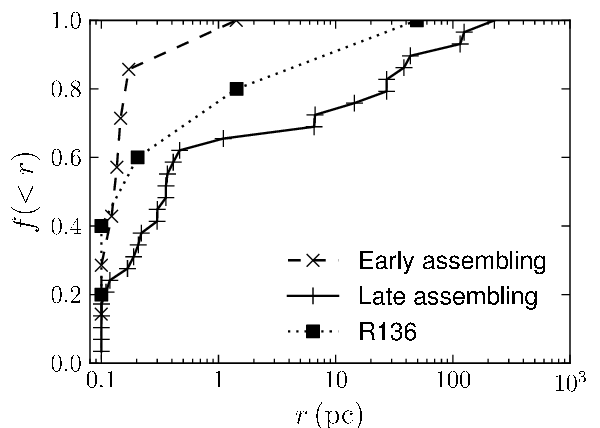}
\caption{Cumulative distribution of PCPs and SCPs as a function of the distance from the cluster center. Dashed and filled curves indicate early and late assembling models, respectively. For the early assembling models, we combined the data from e2k8r1-1, e2k8r1-2, e2k8r2, and e8k8s1 (4 runs), and the average number of PCPs ans SCPs per run is 1.75. For the late assembling models, we combined the data from e2k8r5-1, e2k8r5-2, e2k8r6-1, e2k8r6-2, and e8k8s2 (5 runs), and the average number of PCPs and SCPs is 5.8. Squares indicate the distribution of massive ($>100M_{\odot}$) stars observed in R136 region \citep{2010MNRAS.408..731C,2011A&A...530L..14B}. Since the observation is projected distance, we multiplied them by $\sqrt[]{3}$. For both the simulations and observations, we treat stars within 0.1 pc as at 0.1 pc because the distance is affected by the definition of the cluster center. \label{fig:massive_star}}
\end{center}
\end{figure}

In Figure \ref{fig:massive_star}, we plot the radial distribution
of PCPs and SCPs, which grows to $>100 M_{\odot}$. 
We combine the results from several runs, separating them in the early and 
late assembling cases. 
While all the PCPs and SCPs are located in the cluster core in the early 
assembling case, $\sim40$\% of them are ejected from the clusters or located 
in the outskirts of the cluster ($>10$ pc) in the late assembling case. 
The numbers of PCPs and SCPs per cluster are on average 1.75 and 5.8 for the 
early and late assembling cases, respectively. In Figure 
\ref{fig:massive_star} we also present the cumulative number 
distribution of stars with $>100M_{\odot}$ in the R136 region 
\citep{2010MNRAS.408..731C,2011A&A...530L..14B}. The number of such 
massive stars and their distribution imply that R136 experienced 
some late assembling, and observationally a sub-cluster has been 
found around R136 \citep{2012ApJ...754L..37S}.

\subsection{Central Density of Remnant Clusters}

\begin{figure}
\begin{center}
\includegraphics[width=84mm]{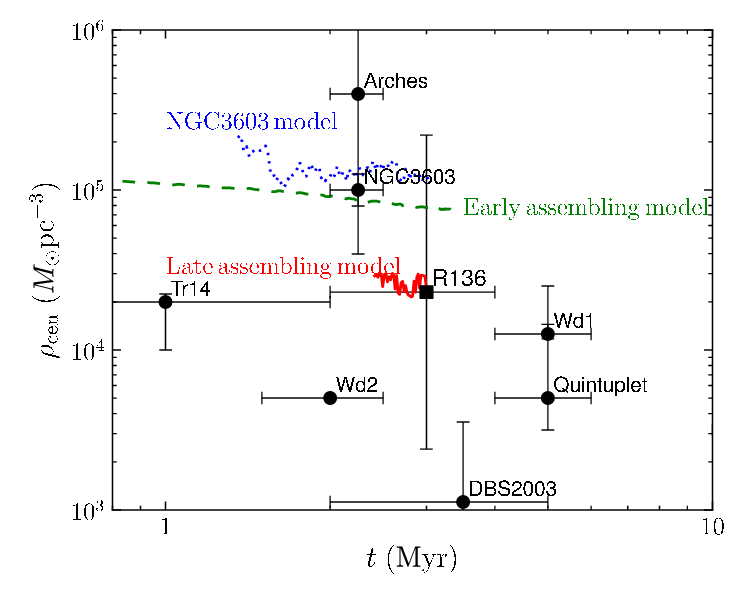}
\caption{The time evolution of the central density ($<0.4$ pc) of remnant clusters 
for assembling models; green-dashed and red-solid curves indicate early and late 
assembling models, respectively. Blue-dotted curves indicate the central density 
($<0.14$ pc) for ensemble models e2kr3. 
Black points with error bars are observed cluster densities from 
\citet{2009A&A...498L..37P}. For R136, the data is from 
\citet{2012MNRAS.424.1372A}. \label{fig:central_density}}
\end{center}
\end{figure}

In Figure \ref{fig:central_density}, we compare the central density of
the remnant clusters with the observed density of young massive
clusters.  Since the core radii and core densities of the simulated
clusters are obtained from rather small number of particles they show
relatively large fluctuations (see Figures \ref{fig:cd} and
\ref{fig:cd_cold}).  We adopted the core radius of NGC3603 of 0.14 pc
for the models e2k4r3, because the mass of these models is comparable
to that of NGC3603 \citep{2008ApJ...675.1319H}.  The total mass of
the models e2k8 is comparable to that of R136, for which we adopted a
core radius of 0.4 pc \citep{2012arXiv1209.3825S} rather than the
observed 0.025 pc \citep{1995ApJ...448..179H,2009ApJ...707.1347A}. We
argue that the observe value is rather strongly affected by the
dynamical evolution of the cluster, and also influenced by the small
number of stars in the core; both tend to cause an under estimate of
the core radius.  In the same figure we also plot the observed
densities of a number of the young clusters listed in
\citep{2009A&A...498L..37P} and \citet{2012MNRAS.424.1372A}.  Those
observed density are not measured in the same way as we determine the
central density in our simulations, although we tried to mimic the
observational technique to measure the core radius as good as
possible. Late assembly models turn out to have a central density
which is an order-of-magnitude lower than in the early assembly
models. Those lower densities very well matches the central density in
the observations of R136. The central density of virialized solo
cluster models evolve quite similarly to the early assembly
simulations.

\section{Summary and Discussion}

We performed $N$-body simulations of solo and ensemble star clusters
and found that ensemble clusters evolve through typically two pathways
depending on their assembling time compared to the core-collapse time
of the sub-clusters.

In the {\em early assembling} case the sub clusters merge before they
experience core collapse individually: $t_{\rm cc}>t_{\rm ens}$.
After merging, their remnant clusters have dynamically mature
characteristics (strongly mass segregated and core collapsed) compared
to solo-clusters.  The early assembling clusters experience mass
segregation and core collapse on the time scale of the sub-clusters,
which is shorter than that of initially large solo clusters; the short
relaxation time of sub-clusters is conserved in the remnant clusters.
This dynamically early evolution results in efficient multiple
collisions of stars and helps the formation of extremely massive 
primary collision products (PCPs)
with $\sim 1000 M_{\odot}$.  The evolution of the early assembling
clusters can be mimicked by solo-clusters when those are born
sub-virial.

In the {\em late assembling} case the sub clusters experience core
collapse before they merge into the larger conglomerate: $t_{\rm
  cc}<t_{\rm ens}$.  The dynamically mature characteristics of the
merger remnant suppresses the growth of massive stars via stellar
collisions. In this case, the sub-clusters experience core collapse
individually and form their own PCPs, but
the maximum mass of the PCPs in the sub-clusters is limited by the
total mass of the sub-clusters. Even after the sub-clusters assemble,
the PCPs stop growing because the central density of the remnant
cluster is already depleted due to the quick dynamical evolution of
the sub-clusters.  Since the PCPs in sub-clusters form massive
binaries, they interact with each other in the remnant clusters. Some
of them (SCPs) collide, but the others are scattered from the cluster
by three-body or binary-binary encounters.  In our simulations, 40\%
of the SCPs are ejected from the cluster or scattered to the outskirts
of the remnant clusters.  The SCPs sometimes escape with a high
velocity ($>30$km/s) and reach $\sim 100$ pc from the cluster within
their life time ($\sim 3$ Myr).  The observed massive high-velocity
stars such as VFTS 682 might be formed in this way (see also Paper 1).

We also investigated the maximum mass of the PCPs and found that in
ensemble clusters, the maximum mass depends on the assembling time of
sub-clusters. In the early assembling models, the maximum mass of the
PCPs is comparable to that of sub-virial solo-clusters.  In the late
assembling models, however, the maximum mass is similar to that of the
solo sub-clusters; the difference is mainly caused by the number of
collisions. In the late assembling models a larger number of SCPs are
ejected from the cluster than in the early assembling case and the
SCPs fail to merge to the PCP.

When the collisions of stars proceed most successfully (in early
assembling and cold solo models), we find that the maximum masses of
the PCPs reach $\sim$2\% of the total mass of the clusters even if we
take into account the high mass-loss rate due to the stellar wind. 
Assuming an R136-like cluster of $\sim
5\times 10^4 M_{\odot}$, the expected maximum mass is $\sim 1000
M_{\odot}$. Such an efficient mass growth might result in the
formation of IMBHs. For lower metalicity, the massive stars are
predicted to collapse directly to IMBHs \citep{2003ApJ...591..288H}.

In late assembling cases, however, a smaller PCP and multiple SCPs 
($100$--$400 M_{\odot}$) are expected to exist inside or around the remnant 
clusters. These stars are in the mass range of 
type Ib/c supernovae (SNe) assuming solar metallicity 
\citep{2003ApJ...591..288H}. 
In recent observations of dense molecular clouds in the central molecular 
zone in the Galactic center, several expanding shells were found, 
and the estimated total kinetic energy of them is $\sim 10^{52}$ erg.
\citep{2007PASJ...59..323T, 2012ApJS..201...14O}. Especially, 
three major shells have a kinetic energy of $\sim 10^{51}$ erg, which
corresponds to a hypernova explosion. 
A young dense massive clusters which is similar to our late-merger 
models might be embedded in this dense molecular cloud.
Furthermore, escaping SCPs will explode up to $\sim 100$ pc 
from the host cluster. 
Actually type Ib/c SNe associate with star forming regions 
\citep{2010MNRAS.407.2660A,2011A&A...530A..95L,2012MNRAS.424.1372A,
2012arXiv1210.1126C}, 
and for example Type Ic SN 2007gr is 
located at $\sim 7$ pc from a young cluster \citep{2008ApJ...672L..99C}.

\section*{Acknowledgments}
The authors thank Jeroen B\'{e}dorf for the Sapppro2 library, Alex
Rimoldi for careful reading of the manuscript, and Masaomi Tanaka for
fruitful discussion.
This work was supported by the Japan Society for the Promotion of 
Science (JSPS) Research Fellowship for Research Abroad, the 
Netherlands Research Council NWO 
(grants \#643.200.503, \#639.073.803 and \#614.061.608), the 
Netherlands Research School for Astronomy (NOVA). 
Numerical computations were carried out on the Cray XT4 at the
Center for Computational Astrophysics (CfCA) of the National 
Astronomical Observatory of Japan and the Little Green Machine 
at Leiden University.

\bibliographystyle{mn}
\bibliography{mn-jour,reference}

\begin{thebibliography}{65}
\expandafter\ifx\csname natexlab\endcsname\relax\def\natexlab#1{#1}\fi

\bibitem[{{Aarseth} \& {Hills}(1972)}]{1972A&A....21..255A}
{Aarseth} S.~J., {Hills} J.~G., 1972, \aap, 21, 255

\bibitem[{{Allison} {et~al.}(2009){Allison}, {Goodwin}, {Parker}, {de Grijs},
  {Portegies Zwart}, \& {Kouwenhoven}}]{2009ApJ...700L..99A}
{Allison} R.~J., {Goodwin} S.~P., {Parker} R.~J., {de Grijs} R., {Portegies
  Zwart} S.~F., {Kouwenhoven} M.~B.~N., 2009, \apjl, 700, L99

\bibitem[{{Andersen} {et~al.}(2009){Andersen}, {Zinnecker}, {Moneti},
  {McCaughrean}, {Brandl}, {Brandner}, {Meylan}, \&
  {Hunter}}]{2009ApJ...707.1347A}
{Andersen} M., {Zinnecker} H., {Moneti} A., {McCaughrean} M.~J., {Brandl} B.,
  {Brandner} W., {Meylan} G., {Hunter} D., 2009, \apj, 707, 1347

\bibitem[{{Anderson} {et~al.}(2010){Anderson}, {Covarrubias}, {James}, {Hamuy},
  \& {Habergham}}]{2010MNRAS.407.2660A}
{Anderson} J.~P., {Covarrubias} R.~A., {James} P.~A., {Hamuy} M., {Habergham}
  S.~M., 2010, \mnras, 407, 2660

\bibitem[{{Anderson} {et~al.}(2012){Anderson}, {Habergham}, {James}, \&
  {Hamuy}}]{2012MNRAS.424.1372A}
{Anderson} J.~P., {Habergham} S.~M., {James} P.~A., {Hamuy} M., 2012, \mnras,
  424, 1372

\bibitem[{{Ardi} {et~al.}(2008){Ardi}, {Baumgardt}, \&
  {Mineshige}}]{2008ApJ...682.1195A}
{Ardi} E., {Baumgardt} H., {Mineshige} S., 2008, \apj, 682, 1195

\bibitem[{{Ascenso} {et~al.}(2007){Ascenso}, {Alves}, {Beletsky}, \&
  {Lago}}]{2007A&A...466..137A}
{Ascenso} J., {Alves} J., {Beletsky} Y., {Lago} M.~T.~V.~T., 2007, \aap, 466,
  137

\bibitem[{{Baumgardt} \& {Klessen}(2011)}]{2011MNRAS.413.1810B}
{Baumgardt} H., {Klessen} R.~S., 2011, \mnras, 413, 1810

\bibitem[{{Belkus} {et~al.}(2007){Belkus}, {Van Bever}, \&
  {Vanbeveren}}]{2007ApJ...659.1576B}
{Belkus} H., {Van Bever} J., {Vanbeveren} D., 2007, \apj, 659, 1576

\bibitem[{{Bestenlehner} {et~al.}(2011){Bestenlehner}, {Vink}, {Gr{\"a}fener},
  {Najarro}, {Evans}, {Bastian}, {Bonanos}, {Bressert}, {Crowther}, {Doran},
  {Friedrich}, {H{\'e}nault-Brunet}, {Herrero}, {de Koter}, {Langer}, {Lennon},
  {Ma{\'{\i}}z Apell{\'a}niz}, {Sana}, {Soszynski}, \&
  {Taylor}}]{2011A&A...530L..14B}
{Bestenlehner} J.~M., {Vink} J.~S., {Gr{\"a}fener} G., {Najarro} F., {Evans}
  C.~J., {Bastian} N., {Bonanos} A.~Z., {Bressert} E., {Crowther} P.~A.,
  {Doran} E., {Friedrich} K., {H{\'e}nault-Brunet} V., {Herrero} A., {de Koter}
  A., {Langer} N., {Lennon} D.~J., {Ma{\'{\i}}z Apell{\'a}niz} J., {Sana} H.,
  {Soszynski} I., {Taylor} W.~D., 2011, \aap, 530, L14+

\bibitem[{{Bonnell} {et~al.}(2011){Bonnell}, {Smith}, {Clark}, \&
  {Bate}}]{2011MNRAS.410.2339B}
{Bonnell} I.~A., {Smith} R.~J., {Clark} P.~C., {Bate} M.~R., 2011, \mnras, 410,
  2339

\bibitem[{{Brandl} {et~al.}(2007){Brandl}, {Portegies Zwart}, {Moffat}, \&
  {Chernoff}}]{2007ASPC..367..629B}
{Brandl} B.~R., {Portegies Zwart} S.~F., {Moffat} A.~F.~J., {Chernoff} D.~F.,
  2007, in Astronomical Society of the Pacific Conference Series, Vol. 367,
  Massive Stars in Interactive Binaries, {N.~St.-Louis \& A.~F.~J.~Moffat},
  ed., p. 629

\bibitem[{{Brandner} {et~al.}(2008){Brandner}, {Clark}, {Stolte}, {Waters},
  {Negueruela}, \& {Goodwin}}]{2008A&A...478..137B}
{Brandner} W., {Clark} J.~S., {Stolte} A., {Waters} R., {Negueruela} I.,
  {Goodwin} S.~P., 2008, \aap, 478, 137

\bibitem[{{Casertano} \& {Hut}(1985)}]{1985ApJ...298...80C}
{Casertano} S., {Hut} P., 1985, \apj, 298, 80

\bibitem[{{Clark} {et~al.}(2005){Clark}, {Negueruela}, {Crowther}, \&
  {Goodwin}}]{2005A&A...434..949C}
{Clark} J.~S., {Negueruela} I., {Crowther} P.~A., {Goodwin} S.~P., 2005, \aap,
  434, 949

\bibitem[{{Crockett} {et~al.}(2008){Crockett}, {Maund}, {Smartt}, {Mattila},
  {Pastorello}, {Smoker}, {Stephens}, {Fynbo}, {Eldridge}, {Danziger}, \&
  {Benn}}]{2008ApJ...672L..99C}
{Crockett} R.~M., {Maund} J.~R., {Smartt} S.~J., {Mattila} S., {Pastorello} A.,
  {Smoker} J., {Stephens} A.~W., {Fynbo} J., {Eldridge} J.~J., {Danziger}
  I.~J., {Benn} C.~R., 2008, \apjl, 672, L99

\bibitem[{{Crowther}(2012)}]{2012arXiv1210.1126C}
{Crowther} P.~A., 2012, ArXiv e-prints

\bibitem[{{Crowther} {et~al.}(2010){Crowther}, {Schnurr}, {Hirschi}, {Yusof},
  {Parker}, {Goodwin}, \& {Kassim}}]{2010MNRAS.408..731C}
{Crowther} P.~A., {Schnurr} O., {Hirschi} R., {Yusof} N., {Parker} R.~J.,
  {Goodwin} S.~P., {Kassim} H.~A., 2010, \mnras, 408, 731

\bibitem[{{Ebisuzaki} {et~al.}(2001){Ebisuzaki}, {Makino}, {Tsuru}, {Funato},
  {Portegies Zwart}, {Hut}, {McMillan}, {Matsushita}, {Matsumoto}, \&
  {Kawabe}}]{2001ApJ...562L..19E}
{Ebisuzaki} T., {Makino} J., {Tsuru} T.~G., {Funato} Y., {Portegies Zwart} S.,
  {Hut} P., {McMillan} S., {Matsushita} S., {Matsumoto} H., {Kawabe} R., 2001,
  \apjl, 562, L19

\bibitem[{{Evans} {et~al.}(2010){Evans}, {Walborn}, {Crowther},
  {H{\'e}nault-Brunet}, {Massa}, {Taylor}, {Howarth}, {Sana}, {Lennon}, \& {van
  Loon}}]{2010ApJ...715L..74E}
{Evans} C.~J., {Walborn} N.~R., {Crowther} P.~A., {H{\'e}nault-Brunet} V.,
  {Massa} D., {Taylor} W.~D., {Howarth} I.~D., {Sana} H., {Lennon} D.~J., {van
  Loon} J.~T., 2010, \apjl, 715, L74

\bibitem[{{Freitag} {et~al.}(2006){Freitag}, {G{\"u}rkan}, \&
  {Rasio}}]{2006MNRAS.368..141F}
{Freitag} M., {G{\"u}rkan} M.~A., {Rasio} F.~A., 2006, \mnras, 368, 141

\bibitem[{{Fujii} {et~al.}(2008){Fujii}, {Iwasawa}, {Funato}, \&
  {Makino}}]{2008ApJ...686.1082F}
{Fujii} M., {Iwasawa} M., {Funato} Y., {Makino} J., 2008, \apj, 686, 1082

\bibitem[{{Fujii} {et~al.}(2009){Fujii}, {Iwasawa}, {Funato}, \&
  {Makino}}]{2009ApJ...695.1421F}
---, 2009, \apj, 695, 1421

\bibitem[{{Fujii} \& {Portegies Zwart}(2011)}]{2011Sci...334.1380F}
{Fujii} M.~S., {Portegies Zwart} S., 2011, Science, 334, 1380

\bibitem[{{Fujii} {et~al.}(2012){Fujii}, {Saitoh}, \& {Portegies
  Zwart}}]{2012ApJ...753...85F}
{Fujii} M.~S., {Saitoh} T.~R., {Portegies Zwart} S.~F., 2012, \apj, 753, 85

\bibitem[{{Gaburov} {et~al.}(2008){Gaburov}, {Gualandris}, \& {Portegies
  Zwart}}]{2008MNRAS.384..376G}
{Gaburov} E., {Gualandris} A., {Portegies Zwart} S., 2008, \mnras, 384, 376

\bibitem[{{Gennaro} {et~al.}(2011){Gennaro}, {Brandner}, {Stolte}, \&
  {Henning}}]{2011MNRAS.412.2469G}
{Gennaro} M., {Brandner} W., {Stolte} A., {Henning} T., 2011, \mnras, 412, 2469

\bibitem[{{Glebbeek} {et~al.}(2009){Glebbeek}, {Gaburov}, {de Mink}, {Pols}, \&
  {Portegies Zwart}}]{2009A&A...497..255G}
{Glebbeek} E., {Gaburov} E., {de Mink} S.~E., {Pols} O.~R., {Portegies Zwart}
  S.~F., 2009, \aap, 497, 255

\bibitem[{{Goswami} {et~al.}(2012){Goswami}, {Umbreit}, {Bierbaum}, \&
  {Rasio}}]{2012ApJ...752...43G}
{Goswami} S., {Umbreit} S., {Bierbaum} M., {Rasio} F.~A., 2012, \apj, 752, 43

\bibitem[{{G{\"u}rkan} {et~al.}(2004){G{\"u}rkan}, {Freitag}, \&
  {Rasio}}]{2004ApJ...604..632G}
{G{\"u}rkan} M.~A., {Freitag} M., {Rasio} F.~A., 2004, \apj, 604, 632

\bibitem[{{Gvaramadze} \& {Gualandris}(2011)}]{2011MNRAS.410..304G}
{Gvaramadze} V.~V., {Gualandris} A., 2011, \mnras, 410, 304

\bibitem[{{Harayama} {et~al.}(2008){Harayama}, {Eisenhauer}, \&
  {Martins}}]{2008ApJ...675.1319H}
{Harayama} Y., {Eisenhauer} F., {Martins} F., 2008, \apj, 675, 1319

\bibitem[{{Heger} {et~al.}(2003){Heger}, {Fryer}, {Woosley}, {Langer}, \&
  {Hartmann}}]{2003ApJ...591..288H}
{Heger} A., {Fryer} C.~L., {Woosley} S.~E., {Langer} N., {Hartmann} D.~H.,
  2003, \apj, 591, 288

\bibitem[{{Heggie} \& {Hut}(2003)}]{2003gmbp.book.....H}
{Heggie} D., {Hut} P., 2003, {The Gravitational Million-Body Problem: A
  Multidisciplinary Approach to Star Cluster Dynamics}, Heggie D. \&~Hut P.,
  ed.

\bibitem[{{Hunter} {et~al.}(1995){Hunter}, {Shaya}, {Holtzman}, {Light},
  {O'Neil}, \& {Lynds}}]{1995ApJ...448..179H}
{Hunter} D.~A., {Shaya} E.~J., {Holtzman} J.~A., {Light} R.~M., {O'Neil} Jr.
  E.~J., {Lynds} R., 1995, \apj, 448, 179

\bibitem[{{Hurley} {et~al.}(2000){Hurley}, {Pols}, \&
  {Tout}}]{2000MNRAS.315..543H}
{Hurley} J.~R., {Pols} O.~R., {Tout} C.~A., 2000, \mnras, 315, 543

\bibitem[{{King}(1966)}]{1966AJ.....71...64K}
{King} I.~R., 1966, \aj, 71, 64

\bibitem[{{Leloudas} {et~al.}(2011){Leloudas}, {Gallazzi}, {Sollerman},
  {Stritzinger}, {Fynbo}, {Hjorth}, {Malesani}, {Micha{\l}owski},
  {Milvang-Jensen}, \& {Smith}}]{2011A&A...530A..95L}
{Leloudas} G., {Gallazzi} A., {Sollerman} J., {Stritzinger} M.~D., {Fynbo}
  J.~P.~U., {Hjorth} J., {Malesani} D., {Micha{\l}owski} M.~J.,
  {Milvang-Jensen} B., {Smith} M., 2011, \aap, 530, A95

\bibitem[{{Mackey} \& {Gilmore}(2003)}]{2003MNRAS.338...85M}
{Mackey} A.~D., {Gilmore} G.~F., 2003, \mnras, 338, 85

\bibitem[{{Massey} \& {Hunter}(1998)}]{1998ApJ...493..180M}
{Massey} P., {Hunter} D.~A., 1998, \apj, 493, 180

\bibitem[{{McMillan} {et~al.}(2007){McMillan}, {Vesperini}, \& {Portegies
  Zwart}}]{2007ApJ...655L..45M}
{McMillan} S.~L.~W., {Vesperini} E., {Portegies Zwart} S.~F., 2007, \apjl, 655,
  L45

\bibitem[{{Mengel} \& {Tacconi-Garman}(2009)}]{2009Ap&SS.324..321M}
{Mengel} S., {Tacconi-Garman} L.~E., 2009, \apss, 324, 321

\bibitem[{{Moeckel} \& {Bonnell}(2009)}]{2009MNRAS.400..657M}
{Moeckel} N., {Bonnell} I.~A., 2009, \mnras, 400, 657

\bibitem[{{Moeckel} \& {Clarke}(2011)}]{2011MNRAS.410.2799M}
{Moeckel} N., {Clarke} C.~J., 2011, \mnras, 410, 2799

\bibitem[{{Nitadori} \& {Makino}(2008)}]{2008NewA...13..498N}
{Nitadori} K., {Makino} J., 2008, New Astronomy, 13, 498

\bibitem[{{Oka} {et~al.}(2012){Oka}, {Onodera}, {Nagai}, {Tanaka}, {Matsumura},
  \& {Kamegai}}]{2012ApJS..201...14O}
{Oka} T., {Onodera} Y., {Nagai} M., {Tanaka} K., {Matsumura} S., {Kamegai} K.,
  2012, \apjs, 201, 14

\bibitem[{{Pauldrach} {et~al.}(2012){Pauldrach}, {Vanbeveren}, \&
  {Hoffmann}}]{2012A&A...538A..75P}
{Pauldrach} A.~W.~A., {Vanbeveren} D., {Hoffmann} T.~L., 2012, \aap, 538, A75

\bibitem[{{Pfalzner}(2009)}]{2009A&A...498L..37P}
{Pfalzner} S., 2009, \aap, 498, L37

\bibitem[{{Portegies Zwart} {et~al.}(2004){Portegies Zwart}, {Baumgardt},
  {Hut}, {Makino}, \& {McMillan}}]{2004Natur.428..724P}
{Portegies Zwart} S.~F., {Baumgardt} H., {Hut} P., {Makino} J., {McMillan}
  S.~L.~W., 2004, \nat, 428, 724

\bibitem[{{Portegies Zwart} {et~al.}(1999){Portegies Zwart}, {Makino},
  {McMillan}, \& {Hut}}]{1999A&A...348..117P}
{Portegies Zwart} S.~F., {Makino} J., {McMillan} S.~L.~W., {Hut} P., 1999,
  \aap, 348, 117

\bibitem[{{Portegies Zwart} \& {McMillan}(2002)}]{2002ApJ...576..899P}
{Portegies Zwart} S.~F., {McMillan} S.~L.~W., 2002, \apj, 576, 899

\bibitem[{{Portegies Zwart} \& {Verbunt}(1996)}]{1996A&A...309..179P}
{Portegies Zwart} S.~F., {Verbunt} F., 1996, \aap, 309, 179

\bibitem[{{Portegies Zwart} \& {Yungelson}(1998)}]{1998A&A...332..173P}
{Portegies Zwart} S.~F., {Yungelson} L.~R., 1998, \aap, 332, 173

\bibitem[{{Rauw} {et~al.}(2007){Rauw}, {Manfroid}, {Gosset}, {Naz{\'e}},
  {Sana}, {De Becker}, {Foellmi}, \& {Moffat}}]{2007A&A...463..981R}
{Rauw} G., {Manfroid} J., {Gosset} E., {Naz{\'e}} Y., {Sana} H., {De Becker}
  M., {Foellmi} C., {Moffat} A.~F.~J., 2007, \aap, 463, 981

\bibitem[{{Roman-Lopes} {et~al.}(2011){Roman-Lopes}, {Barba}, \&
  {Morrell}}]{2011MNRAS.416..501R}
{Roman-Lopes} A., {Barba} R.~H., {Morrell} N.~I., 2011, \mnras, 416, 501

\bibitem[{{Sabbi} {et~al.}(2012){Sabbi}, {Lennon}, {Gieles}, {de Mink},
  {Walborn}, {Anderson}, {Bellini}, {Panagia}, {van der Marel}, \& {Ma{\'{\i}}z
  Apell{\'a}niz}}]{2012ApJ...754L..37S}
{Sabbi} E., {Lennon} D.~J., {Gieles} M., {de Mink} S.~E., {Walborn} N.~R.,
  {Anderson} J., {Bellini} A., {Panagia} N., {van der Marel} R., {Ma{\'{\i}}z
  Apell{\'a}niz} J., 2012, \apjl, 754, L37

\bibitem[{{Saitoh} {et~al.}(2011){Saitoh}, {Daisaka}, {Kokubo}, {Makino},
  {Okamoto}, {Tomisaka}, {Wada}, \& {Yoshida}}]{2011IAUS..270..483S}
{Saitoh} T.~R., {Daisaka} H., {Kokubo} E., {Makino} J., {Okamoto} T.,
  {Tomisaka} K., {Wada} K., {Yoshida} N., 2011, in IAU Symposium, Vol. 270, IAU
  Symposium, {J.~Alves, B.~G.~Elmegreen, J.~M.~Girart, \& V.~Trimble}, ed., pp.
  483--486

\bibitem[{{Salpeter}(1955)}]{1955ApJ...121..161S}
{Salpeter} E.~E., 1955, \apj, 121, 161

\bibitem[{{Selman} \& {Melnick}(2012)}]{2012arXiv1209.3825S}
{Selman} F.~J., {Melnick} J., 2012, ArXiv e-prints

\bibitem[{{Smith} {et~al.}(2011){Smith}, {Slater}, {Fellhauer}, {Goodwin}, \&
  {Assmann}}]{2011MNRAS.416..383S}
{Smith} R., {Slater} R., {Fellhauer} M., {Goodwin} S., {Assmann} P., 2011,
  \mnras, 416, 383

\bibitem[{{Stolte} {et~al.}(2006){Stolte}, {Brandner}, {Brandl}, \&
  {Zinnecker}}]{2006AJ....132..253S}
{Stolte} A., {Brandner} W., {Brandl} B., {Zinnecker} H., 2006, \aj, 132, 253

\bibitem[{{Tanaka} {et~al.}(2007){Tanaka}, {Kamegai}, {Nagai}, \&
  {Oka}}]{2007PASJ...59..323T}
{Tanaka} K., {Kamegai} K., {Nagai} M., {Oka} T., 2007, \pasj, 59, 323

\bibitem[{{Toonen} {et~al.}(2012){Toonen}, {Nelemans}, \& {Portegies
  Zwart}}]{2012A&A...546A..70T}
{Toonen} S., {Nelemans} G., {Portegies Zwart} S., 2012, \aap, 546, A70

\bibitem[{{Vesperini} {et~al.}(2009){Vesperini}, {McMillan}, \& {Portegies
  Zwart}}]{2009Ap&SS.324..277V}
{Vesperini} E., {McMillan} S., {Portegies Zwart} S., 2009, \apss, 324, 277

\bibitem[{{Yu} {et~al.}(2011){Yu}, {de Grijs}, \& {Chen}}]{2011ApJ...732...16Y}
{Yu} J., {de Grijs} R., {Chen} L., 2011, \apj, 732, 16

\end{thebibliography}

\label{lastpage}

\end{document}